\documentclass[conference]{IEEEtran}

\usepackage{algorithm}
\usepackage{listings}
\usepackage{array}
\usepackage{amssymb}
\usepackage{balance}
\usepackage{graphicx}
\usepackage{pgfplots}
\usepackage{epstopdf}
\usepackage{times}
\usepackage{multirow}
\usepackage{subfig}
\usepackage{hyperref}
\usepackage{amsmath}
\usepackage{listings}
\usepackage[font={small}]{caption}
\ifCLASSINFOpdf
\else
\fi

\begin{document}
\title{How the Sando Search Tool Recommends Queries}

\author{\IEEEauthorblockN{Xi Ge\IEEEauthorrefmark{1}, David Shepherd\IEEEauthorrefmark{2}, Kostadin Damevski\IEEEauthorrefmark{3}, Emerson Murphy-Hill\IEEEauthorrefmark{1}}
\IEEEauthorblockA{\IEEEauthorrefmark{1}NC State University, Raleigh, NC, USA\\
xge@ncsu.edu, emerson@csc.ncsu.edu}
\IEEEauthorblockA{\IEEEauthorrefmark{2}ABB Inc, Raleigh, NC, USA\\
david.shepherd@us.abb.com}
\IEEEauthorblockA{\IEEEauthorrefmark{3}Virginia State University, Petersburg,
VA, USA\\
kdamevski@vsu.edu}}


%Telephone: (800) 555--1212, Fax: (888) 555--1212}


\maketitle

\begin{abstract}
Developers spend a significant amount of time searching their local codebase.
To help them search efficiently, researchers have proposed novel tools that
apply state-of-the-art information retrieval algorithms to retrieve relevant
code snippets from the local codebase. However, these tools still rely on the
developer to craft an effective query, which requires that the developer is
familiar with the terms contained in the related code snippets. Our empirical
data from a state-of-the-art local code search tool, called Sando, suggests that 
developers are sometimes unacquainted with their local codebase. In order 
to bridge the gap between developers and their ever-increasing local codebase, 
in this paper we demonstrate the recommendation techniques integrated in Sando.
\end{abstract}

\section{Introduction}
\label{sec:introduction}
Software is hard to maintain. One cause for the difficulty lies in the
increasing complexity of software systems. To accomplish a software maitenance
task, a developer needs to explore the information space of a software system
and comprehend all relevant parts of the codebase. This exploration is both
tedious and time-consuming. According to Singer et al., developers spend
over $40\%$ of their time in navigating, searching and reading source
code~\cite{Singer}.

Due to the overwhelming complexity of software systems, developers often start
maintenance tasks by searching for a starting point in the local codebase. A
recent study conducted by Ko et al. indicates that most ($9$ out of
$12$) software maintenance tasks start with local code search~\cite{Ko}.

To help developers search their codebase, researchers have proposed several
local code search tools that are integrated into popular IDEs. For instance,
Sando enhances the Visual Studio IDE and InstaSearch is aimed for Eclipse
developers~\cite{Insta, Shepherd}. Both of these search tools significantly
improve upon the built-in search tools available in these IDEs in the following
aspects: (1) they retrieve different levels of software entities, such as
fields, methods, and classes, as search results, instead of retrieving lines of
text; (2) they rank the search results according to their relevance to the given
search query; and (3) these tools support multi-word searches, that is, querying
by space-separated terms and retrieving software entities containing any of
these terms.

% However, all search tools have a common assumption that prevents them
% from being effective at any situations.  The local code search tools
% assume that as a developer, she always know the text contained by
% the code snippets she intends, such as variable or method
% names. Therefore, the developer can query with the text.

Regardless of their sophistication, all search tools require developers issuing
both specific and relevant queries to produce useful results. However, the size
of the codebase under search as well as the vocabulary mismatch problem, where
a domain concept is expressed differently in the code, often make this
requirement unreasonable. Empirical data collected from Sando usage in the field
further substantiates this conjecture. Specifically, about $20\%$ of all Sando
queries fail, returning no results at all, which suggests developers in the
field are facing difficulties in writing effective queries.

To solve this problem, we improve Sando by a set of query recommendation
techniques. These techniques fall into two categories: pre-search
recommendations and post-search recommendations. The former recommends queries
before a developer starts searching and the later recommends queries after a
developer's search fails, that is, when the search returns no results. Below, we
present how these recommendation techniques assist developers to compose queries
in a semi-automatic fashion that are better at finding the relevant code
snippets.

\section{Approach}
\begin{figure*}[th!]
  \centering
      \includegraphics[width=.9\linewidth]{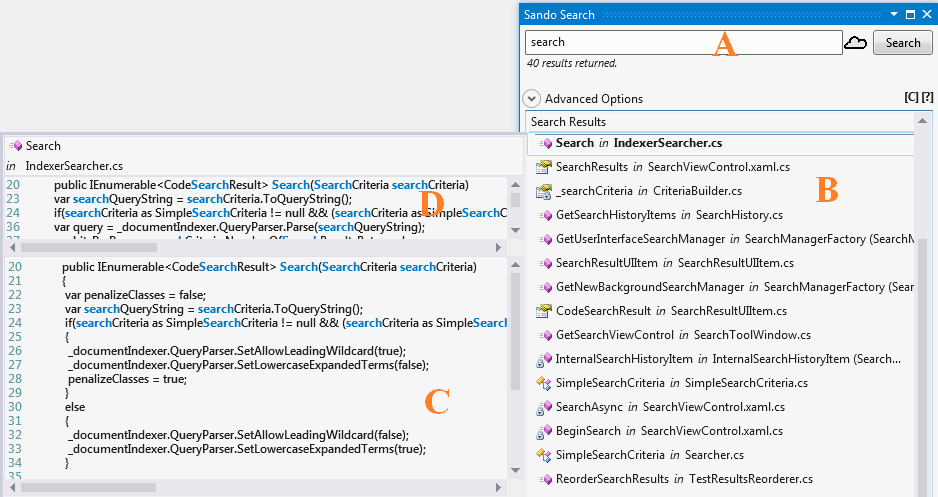}
  \caption{{\footnotesize Sando screenshot after retrieving search results.}}
  \label{fig:sando}
\end{figure*}

Before detailing the recommendation techniques, we first briefly introduce the
local code search tool called Sando, the platform on which our query
recommendation techniques are implemented and evaluated. Sando is a
state-of-the-art code search tool for Visual Studio developers~\cite{Shepherd},
which up to this point has been downloaded over three thousand times. According
to our usage data collected from Sando users, roughly forty developers use Sando
on a daily basis. Figure~\ref{fig:sando} presents the user interface of Sando.
Sando supports multiple programming languages including C\#, C++, C and XML.

After installing Sando, a developer can issue queries in the search box at A,
press the button on the right, and be presented by search results at B. Next,
the developer can examine the search results in one of two ways:
by single clicking on a search result or by double clicking the search result. A
single click triggers a pop-up menu that summarizes the search result, as
illustrated by C and D in Figure~\ref{fig:sando}, where C shows the
entire software entity, which can be the declaration of a class, a method, or a
field; and D shows the lines of code containing the search query; the developer
can also double click the search result to open the containing file in the Visual Studio editor.

\subsection{Components}
To recommend queries, our recommendation techniques rely on five different data
source components, namely the local dictionary, the term co-occurrence matrix, 
the verb-direct-object pairs, the software engineering thesaurus, and the
general English thesaurus, which we will explain next.

\subsubsection{Local Dictionary}
\label{sec:local}
The first component used by our recommendation techniques is the local
dictionary of the codebase under search. Specifically, the local dictionary
contains terms that appear in the codebase at least once. To construct the
dictionary, we reuse the indexing performed by the Sando search engine. The
indexing process breaks each software entity from the codebase into terms;
later, after a developer issues a search query, the search engine retrieves the
software entities whose indexed terms match with the given query as search
results. Taking the following code snippet as an example, the Sando search
engine indexes the method \texttt{Perform} to the terms of ``perform'',
``output'', ``func'', ``invoke'', ``input'', ``finished'', ``event'', and
``finishedevent''. In addition to indexing raw identifiers, Sando indexes the
terms that result from splitting the identifiers that are in camel case or
underscore delimited format.

\small
\begin{verbatim}
    void Perform()
    {
        var output = func.Invoke(input);
        if(FinishedEvent != null)
            FinishedEvent(this, output);
    }
\end{verbatim}

\normalsize

Sando performs either a entire indexing or an incremental indexing: entire
indexing, performed when a developer opens a new project that Sando has no
cached index, traverses and collects terms for the entire project; incremental
indexing, on the other hand, monitors the changed part of a cached project and
re-indexes that part. Indexing the entire project of $10K$ LOC takes about $30$
seconds to finish; incrementally indexing an updated C\# file takes about $30$
milliseconds.

The local dictionary consists of all the terms collected from indexing without
redundancy. To facilitate searching this dictionary, we construct a binary
search tree on these terms. Taking the codebase of Sando as an example, the
local dictionary contains about two thousand different terms. Finding a given
term in the local dictionary costs trivial time to finish. Using the local
dictionary, our recommendation technique can ensure that the recommended queries
actually appear in the local codebase under search.

\subsubsection{Term Co-occurrence Matrix}
In addition to the local dictionary, we maintain the collected terms from the
local codebase through a co-occurrence matrix. The matrix has an equal number of
rows and columns; each column or row represents one term appearing in the local
dictionary. Each element in the matrix saves the count of two terms, as
represented by the row and the column that occur together in the codebase. For
instance, the element at [red, blue] is the count of occurrences of ``red'' and
``blue'' together. By appearing together, we mean these two terms appear in the
same software entity. Again taking the method \texttt{Perform} in
Section~\ref{sec:local}as an example, because the method is an independent
software entity, any two of the indexed eight terms appear together.

Literally keeping the full co-occurrence matrix in memory is inefficient due
to the observation that the matrix is usually sparse. To improve the memory-efficiency,
our technique keep the matrix by using Yale format, a data structure that
maintains a sparse matrix using three rows of integers~\cite{Schultz}.

\subsubsection{Verb-direct-object Pairs}
Many code snippets conceptually correspond to performing certain actions on
certain objects, such as ``open file'', ``close stream'' and ``create
instance''. Based on this observation, Fry et al. proposed a text mining
technique that collects these concepts, or verb-direct-object pairs, from a
given codebase~\cite{Fry}.
Applying this technique, Sando mines the verb-direct-object pairs from the code
base under search, and caches them for recommending either the verb or the
object when the developer queries the other part. For instance, supposing Sando
mined the verb-direct-object pair ``open file'' from the codebase under search.
If the developer queries ``open'', Sando recommends ``file'' to the search
query. Similarly, if the developer queries ``file'', Sando recommends ``open''
to the original query.

\subsubsection{Software Engineering Thesaurus}
Sando's recommendation techniques also tries to solve the vocabulary mismatch
problem, which happens when a developer queries terms that do not appear in the
codebase, however semantically relate to some terms that do. For instance, if
the codebase under search uses ``retrieval'' consistently, then the developer's
querying of ``search'' are unlikely to return useful results. To solve this
problem, we applied a thesaurus-based technique. Taking a term as input, the
thesauri return synonyms that appear in the codebase as recommended queries.
Only using general English thesaurus is not enough because software development
has developed many field-specific synonyms, such as ``instantiate'' to
``create'' and ``update'' to ``refresh''. Therefore, we apply the work of Gupta 
et al. that mined the source code of open software projects, generating $1724$
pairs of related terms. Among these pairs, about $91\%$ are
field-specific~\cite{Gupta}.

\subsubsection{General English Thesaurus}
In addition to the field-specific thesaurus, we also include a general English
thesaurus to help the developer when the field-specific thesaurus fails. We
derive the general English thesaurus from WordNet, which is database of English
words with the relationship between them~\cite{Miller}. Keeping the entire
WordNet in memory is costly; hence we only include the top $100k$ most frequent
terms in the English language.

\subsection{Pre-search Recommendations}
Based on the aforementioned components, we next detail how Sando computes and
presents the recommended queries. The first category of our recommendation
techniques assists developers before they query. These pre-search
recommendations allow developers to select terms that are either frequent in the
local codebase or closely related to the input. More precisely, the recommended
queries originate from the following three sources: (1) the verb-direct-object
pairs, (2) the identifiers, and (3) the co-occurring terms.

\begin{figure}[t]
  \centering
	\subfloat[][{\footnotesize Recommending verb-direct-object pairs.}]{
    \includegraphics[width=0.9\linewidth]{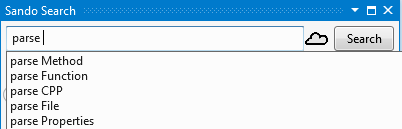}
    \label{fig:verbDO}}
    
     \subfloat[][{\footnotesize Recommending identifiers.}]{
    \includegraphics[width=0.9\linewidth]{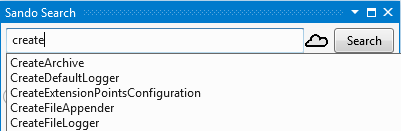}
    \label{fig:identifiers}}
   
   \subfloat[][{\footnotesize Frequently co-occurring terms.}]{
    \includegraphics[width=0.9\linewidth]{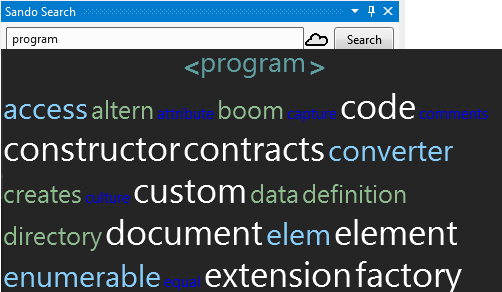}
    \label{fig:tagcloud}}
 \caption{{\footnotesize UI of pre-search recommendations.}}
\label{fig:pre-search}
\end{figure}

When the developer inputs a verb or a noun in the search box, a drop-down menu
presents the verb-direct-object pairs which contain the term given in the search
box. For instance, as illustrate in Figure~\ref{fig:verbDO},  when the developer
inputs ``parse'' to the search box, the drop-down menu lists pairs such as
``parse file'' and ``parse method''.

When the developer inputs the prefix of cached identifiers, the drop-down menu
presents the identifiers starting with the given prefix. For instance, the
drop-down menu in Figure~\ref{fig:identifiers} shows identifiers starting with
``create'', which is the developer's input in the search box. These identifiers
could be method names, class names, and field names. Sando retrieves these
identifiers from the local dictionary.

The third source of the recommended terms is the co-occurrence matrix. After the
developer inputs a term in the search box, she can view the terms that co-occur
with the term. Different from the aforementioned two sources, Sando presents the
co-occurring terms through a tag cloud. The size of a term in the tag cloud
indicates the comparative co-occurrence count. The bigger the font size of a
term, the more frequent the term co-occurs with the term in the search box.

For instance, after the developer inputs ``program'' in the search box and
clicks the cloud button near the search box, a tag cloud appears as illustrated
in Figure~\ref{fig:tagcloud}. From the tag cloud, the developer can infer that
``code'' appears with ``program'' more frequently than ``data'' does.
Furthermore, each term in the tag cloud is a hyper link, the click on which adds
the term to the original query. The reason for using the tag cloud instead of
the drop-down menu is that the co-occurring terms significantly outnumber the
identifiers and the verb-direct-object pairs; presenting the co-occurring terms
through the drop-down menu may be inconvenient for the developer to browse.

\subsection{Post-search Recommendations}
The second category of the recommendation techniques helps developers when their
manual queries fail, which happens when the queries contain terms not appearing
in the local codebase. The post-search recommendation assumes that the failure
is either due to the vocabulary mismatch problem or to misspelling. Hence, the
recommended terms are either the synonyms to the original terms or the original
terms with corrected typos. Figure~\ref{fig:post} depicts the UI of the
post-search recommendations after searching ``choice'' fails.

For a query leading to no search results, the recommendation technique first
pre-processes the query by splitting it. Using the terms in the local
dictionary, we greedily extract the terms of the query that appear in the local
codebase. Our recommendation technique does not recommend replacements for these
terms. For the remaining terms that do not appear in the local dictionary, the
post-search recommendation goes through the following steps to compute the
recommended replacements:

\begin{figure}[t!]
  \centering
      \includegraphics[width=0.9\linewidth]{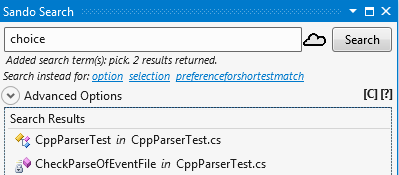}
  \caption{{\footnotesize Post-search recommendations.}}
  \label{fig:post}
\end{figure}

{\bf Step 1:} For a term that does not appear in the local dictionary, Sando
first queries the Software Engineering thesaurus to find the synonyms of the term.
After finding several synonyms, Sando recommends them to the developer. Sando
filters the the recommended sysnonyms with the local dictionary
to ensure that every recommended term appears in the local codebase at least once.

{\bf Step 2:} When finding no field-specific synonyms, our technique queries the
general English dictionary to find synonyms. After finding several synonyms, Sando
recommends them to the developer. Similar with Step 1, Sando also filters these
sysnonmys.

{\bf Step 3:} If neither thesauri contains the synonyms of the given term, Sando
considers the term a typo. Therefore, Sando corrects the typo by using the terms
in the local dictionary.

\section{Related Work}
A large body of existing research relates to ours. We briefly summarize them
according to two aspects: search tools and recommendation techniques.

{\bf Search Tools.} Similar to Sando, researchers have proposed multiple tools
allowing developers to efficiently access the enormous information space. For
instance, Grechanik et al. proposed a search engine called Exemplar that
retrieves function-level code elements from the web and also visualizes
them~\cite{Grechanik}. Bazrafshan et al. presented a search tool that can
retrieve code across the boundary of versions and branches in the source control
system~\cite{Bazrafshan}. Differing from these tools, Sando searches code
snippets from the local codebase.

{\bf Recommender Systems.} Recommender systems play an important role in the
software engineering research. For instance, Thummalapenta and Xie presented
PARSEWeb that recommends method invocations sequence that bridges the starting
object with the destination object~\cite{Thummalapenta}. Ankolekar et al.
proposed Dhruv to recommend reusable artifacts to open source
developers~\cite{Ankolekar}. Differing from these techniques, the recommendation
techniques adopted by Sando assist developers in composing promising queries.

\section{Conclusion}
Facing an ever-increasing codebase, developers often start tasks with searching
their local code. However, even the state-of-the-art code search tools require
developers' recalling the text of the intended code snippets. To further bridge
the gap between developers and their intended code snippets, we demonstrate a
set of recommendation techniques integrated into Sando. Issuing recommendations
both before and after a developer's search, our technique potentially mitigates
developers' cognitive burden when using local code search tools.

\section*{Acknowledgment} This research was conducted during the first author's
internship at ABB Research. We thank the help from Vinay Augustine, Patrick
Francis, and Will Snipes. We also thank the comments from the Development
Liberation Front group members Titus Barik, Michael Bazik, Brittany Johnson,
Kevin Lubick, John Majikes, Yoonki Song and Jim Witschey.

\end{document}